\documentclass[journal=jpccck,manuscript=article,layout=traditional]
{achemso}
\usepackage[version=3]{mhchem} 
\usepackage[]{hyperref}
\usepackage[nameinlink,noabbrev]{cleveref}
\usepackage{siunitx}
\usepackage{xcolor}

\author{Viren Tyagi}
    \affiliation{Materials Simulation \& Modelling, Department of Applied Physics and Science Education, Eindhoven University of Technology, 5600 MB, Eindhoven, The Netherlands}
    \alsoaffiliation{Center for Computational Energy Research, Department of Applied Physics and Science Education, Eindhoven University of Technology, 5600 MB, Eindhoven, The Netherlands}
\author{Mike Pols}
    \affiliation{Materials Simulation \& Modelling, Department of Applied Physics and Science Education, Eindhoven University of Technology, 5600 MB, Eindhoven, The Netherlands}
    \alsoaffiliation{Center for Computational Energy Research, Department of Applied Physics and Science Education, Eindhoven University of Technology, 5600 MB, Eindhoven, The Netherlands}
\author{Geert Brocks}
    \affiliation{Materials Simulation \& Modelling, Department of Applied Physics and Science Education, Eindhoven University of Technology, 5600 MB, Eindhoven, The Netherlands}
    \alsoaffiliation{Center for Computational Energy Research, Department of Applied Physics and Science Education, Eindhoven University of Technology, 5600 MB, Eindhoven, The Netherlands}
    \alsoaffiliation{Computational Chemical Physics, Faculty of Science and Technology and MESA+ Institute for Nanotechnology, University of Twente, 7500 AE, Enschede, The Netherlands}
\author{Shuxia Tao}
    \affiliation{Materials Simulation \& Modelling, Department of Applied Physics and Science Education, Eindhoven University of Technology, 5600 MB, Eindhoven, The Netherlands}
    \alsoaffiliation{Center for Computational Energy Research, Department of Applied Physics and Science Education, Eindhoven University of Technology, 5600 MB, Eindhoven, The Netherlands}
    \email{s.x.tao@tue.nl}

\title{Tracing Ion Migration in Halide Perovskites with Machine Learned Force Fields}


\begin{document}


\begin{abstract}

Halide perovskite optoelectronic devices suffer from chemical degradation and current-voltage hysteresis induced by migration of highly mobile charged defects. Atomic scale molecular dynamics simulations can capture the motion of these ionic defects, but classical force fields are too inflexible to describe their dynamical charge states. Using \ce{CsPbI3} as a case study, we {train} machine learned force fields from density functional theory calculations and study the diffusion of charged halide interstitial and vacancy defects in bulk \ce{CsPbI3}. We find that negative iodide interstitials and positive iodide vacancies, the most stable charge states for their respective defect type, migrate at similar rates at room temperature. Neutral interstitials are faster, but neutral vacancies are one order of magnitude slower. Oppositely charged interstitials and vacancies, as they can occur in device operation or reverse bias conditions, are significantly slower and can be considered relatively immobile.

\end{abstract}


\section{Introduction}
Halide perovskites are becoming prominent in many optoelectronic applications, including solar cells\cite{Liang2023}, light emitting diodes (LEDs)\cite{Lin2018}, and photodetectors\cite{Liu2019}. These materials have an \ce{AMX3} chemical formula, where \ce{A} is a monovalent inorganic or organic cation (\ce{Cs^{+}}, methylammonium \ce{CH_{3}NH_{3}^{+}}, or formamidimium \ce{CH(NH_{2}){_{2}}^{+}}), \ce{M} is a divalent metal cation (\ce{Pb^{2+}}, or \ce{Sn^{2+}}), and \ce{X} is a monovalent halide anion (\ce{I^{-}}, \ce{Br^{-}}, or \ce{Cl^{-}}). Halide perovskites are relatively soft materials. They inherently have a high concentration of intrinsic defects\cite{Wang2018}, which are also quite mobile\cite{ball2016defects}. 

The migration of these defects interferes with device performance. For example, the accumulation of charged defects at the perovskite-electrode interfaces is suggested to cause hysteresis in the current-voltage (I-V) characteristics of these devices\cite{Li2017}. Migration of defects also leads to the degradation of materials and interfaces. During device operation, defects can trigger redox and chemical decomposition reactions\cite{yu2023effect}.
Such effects may negatively impact the optoelectronic properties of the materials, consequently degrading device performance, which is detrimental to the commercialization of perovskite-based optoelectronic devices\cite{chen2022}.

Experimentally, defects are typically characterized through the effects they have on the (thermo)electronic responses of a device \cite{senocrate2017nature,senocrate2019solid,futscher2019quantification,Reichert2020,Ni2022,tammireddy2021temperature,ghasemi2023multiscale} Depending on the specific experimental techniques used, it may be possible to assess the charge states and/or the energy levels of the defects present. However, their chemical composition or atomistic structure remains elusive. From general thermodynamic considerations, it is more likely that defects in bulk materials consist of point defects, i.e., single ion vacancies or interstitials, rather than extended or compound defects \cite{xue2023compound}, which is why both experiment and theory focus on point defects. 

Regarding cation defects, no experimental study seems to indicate the presence of \ce{Pb}-related point defects, or at least, no electronically active, or mobile ones \cite{senocrate2017nature}. Concerning \ce{A} cation defects opinions are more divided, with some studies suggesting the presence of mobile \ce{MA} interstitials\cite{futscher2019quantification,Reichert2020} or vacancies\cite{Reichert2020} in \ce{MAPbI_{3}}, and others finding no evidence for that \cite{senocrate2017nature,tammireddy2021temperature}. In contrast, anion point defects, i.e., halide vacancies and interstitials, are generally considered the dominant mobile species\cite{senocrate2019solid}. Whereas some studies insist on the importance of iodide vacancies in lead iodide perovskites \cite{senocrate2017nature,tammireddy2021temperature}, others instead focus on iodide interstitials \cite{futscher2019quantification,Reichert2020,Ni2022,thiesbrummel2024ion}  

While it is difficult to assess the chemical and atomistic structure of mobile point defects from experiments, atomistic modeling may help to obtain microscopic understanding. { In soft materials such as metal halide perovskites, mobile defects sample a large portion of configuration space. The basic computational technique for assessing diffusion barriers, transition state theory (TST), only samples a small number of migration paths, which at times results in a considerable spread in the numerical values for the barriers, depending on which paths are chosen. Molecular dynamics (MD)simulations are a less biased tool, and have obtained a boost since machine-learned force fields (MLFFs) acquired the accuracy of first-principles calculations \cite{jinnouchi2019phase,Jinnouchi2019}}.

So far, { MLFF MD simulations studying} the motion of defects have focused on neutral defects \cite{balestra2020efficient,pols2023}. In contrast, the defects characterized in the experiments cited above are charged. Indeed, from first-principles calculations, it follows that under equilibrium conditions, halide interstitials and vacancies are negatively and positively charged, respectively\cite{meggiolaro2018,meggiolaro2018first,Xue2021first,Zhang2022,Xue2022}. Moreover, under non-equilibrium conditions, { as they occur under device operating conditions, these defects can change their charge state, where lowering the (quasi) Fermi level can cause halide interstitials to eventually become positively charged \cite{motti2019controlling,Ni2022}, and raising the Fermi level can induce a negative charge on halide vacancies.}

To describe the motion of iodide defects, one has to deal with different charge states { as they occur under different (quasi) Fermi levels}, and even include the possibility that a charge state changes along the defect migration path, as the position of the defect level that traps the charge depends on the local environment of the defect. On the one hand, it is very difficult to capture such elements of charge-dependent migration in a classical force field. On the other hand, quantum mechanics-based \textit{ab-initio} molecular dynamics (AIMD) methods { do this automatically as they incorporate the electrons, but AIMD is} computationally too expensive to reach the required system sizes and timescales for realistic simulations. This is where machine learned force fields (MLFF), on-the-fly trained using AIMD, offer a promising alternative.

In this study, using \ce{CsPbI3} as a model system, we train accurate machine learned force fields (MLFF) on-the-fly using density functional theory (DFT) calculations for different charge { environments} of halide interstitial and vacancy defects. The accuracy of each MLFF is validated by comparing it with DFT-calculated energies, forces, and the energy barriers of typical migration paths. Subsequently, we conduct long-timescale molecular dynamics (MD) simulations at various temperatures to investigate the diffusion behavior of these defects. Our findings indicate that diffusion coefficients and activation barriers of both defect types (vacancy and interstitial) are significantly impacted by their charge environments, with the evolution of structural geometries along the migration path playing a crucial role.   

Similar techniques have been used to study the librational motion of \ce{MA} cations in \ce{MAPbX_{3}}\cite{Bokdam2021exploring}, and \ce{Cs} cation rattling in \ce{CsPbBr_{3}} \cite{lahnsteiner2022anharmonic}, for instance. Here, we use it to trace the motion of iodide point defects in \ce{CsPbI_{3}}. Mean-squared displacements and structural geometries were analyzed to provide atomic-scale insight into the migration behavior of the ions.

The structures of all defect systems were optimized using the Vienna Ab-Initio Simulation Package (VASP)\cite{Kresse1996} with the PBE-D3-BJ exchange-correlation functional\cite{Perdew1996,Grimme2011}. Following structure optimization, the force fields were trained in  VASP, where the training structures were sampled from short-timescale MD runs using Bayesian inference \cite{jinnouchi2019phase,Jinnouchi2019}. A combination of a two-body radial descriptor and a three-body angular descriptor, both of similar forms to the smooth overlap of atomic positions (SOAP) \cite{bartok2013,jinnouchi2020} descriptor, was used to represent the local chemical environments. A variant of Gaussian approximation potentials (GAP), trained on energies, forces, and stress tensors from DFT calculations \cite{bartok2010,jinnouchi2019phase,Jinnouchi2019}, was used to generate the force fields. The latter were then used to perform MD runs in VASP.

\section{Results}
\begin{figure*}[htbp!]
    \includegraphics{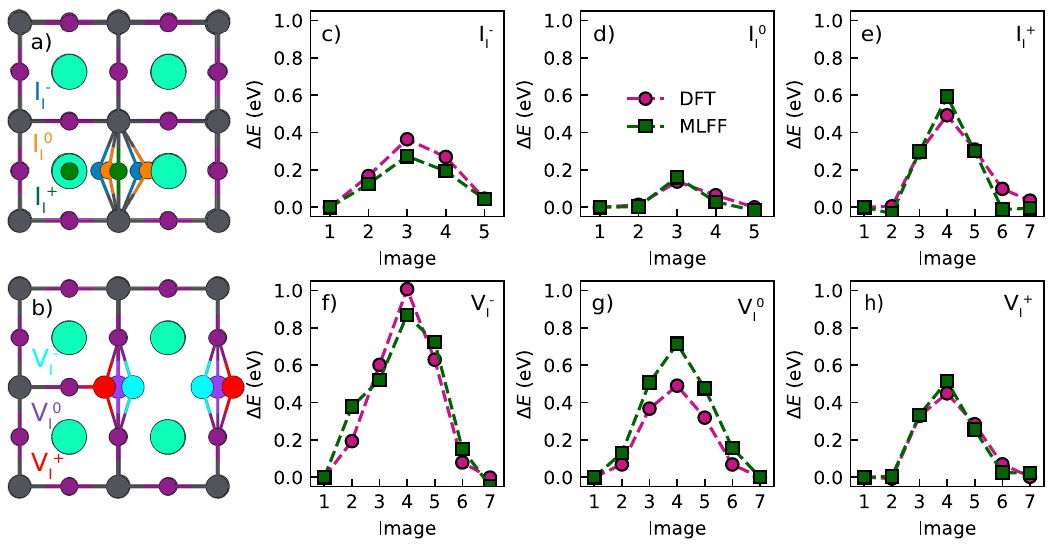}
    \caption{{Schematic structures of (a) the iodide interstitial and (b) the iodide vacancy in \ce{CsPbI_{3}} in their different charge states. (c-h) Energies along the NEB migration paths, calculated using DFT and the MLFFs. The points are the calculated values, and the lines guide the eye; the energy of the minima is set to 0.}}
    \label{fig:figure1}
\end{figure*}

We start by performing DFT calculations to optimize the structures of iodide interstitials and vacancies in three different charge {states}. They include the most stable intrinsic point defects, i.e., the negatively charged iodide interstitial ($\mathrm{I_{I}^{-}}$), and the positively charged iodide vacancy ($\mathrm{V_{I}^{+}}$) in \ce{CsPbI3}. By changing the number of electrons in the supercell, other charge states of the iodide interstitial ($\mathrm{I_{I}^{0}}$ and $\mathrm{I_{I}^{+}}$) and the vacancy ($\mathrm{V_{I}^{0}}$ and $\mathrm{V_{I}^{-}}$) are created. We observe notable local structural changes for both iodide interstitial (Figure~\ref{fig:figure1}a) and iodide vacancy once the charge state changes (Figure~\ref{fig:figure1}b), in agreement with previous work \cite{meggiolaro2018,meggiolaro2018first,Xue2021first,Zhang2022,Xue2022}. 

Band structure calculations are performed to monitor the shift in Fermi level and charge state. These calculations reveal that a decrease of the number of electrons (decrease of the Fermi level) is consistent with $\mathrm{I_{I}^{-}}$ capturing one or two holes, { and becoming $\mathrm{I_{I}^{0}}$ and $\mathrm{I_{I}^{+}}$, respectively,} whereas increasing the number of electrons (increasing the Fermi level) leads to $\mathrm{V_{I}^{+}}$ capturing one or two electrons, { and becoming $\mathrm{V_{I}^{0}}$ and $\mathrm{V_{I}^{-}}$, respectively}. The DFT parameters used for the structure optimizations {and the optimized defect geometries}, as well as the band structures of all defective supercells are given in SI Note 1. { The analysis of the band structures is consistent with the distributions of the excess charge in the defective supercell, as discussed in SI note 2  \cite{Meggiolaro2020tin}.}

Following structure optimization, we train different MLFFs for all six defect systems at different temperatures over a range from 600K to 750K. The training runs for the defect systems are performed using $2\times2\times2$ cubic supercells (8 units of \ce{CsPbI3}) with one iodide point defect. { To check the impact of spin-orbit coupling (SOC), the optimized geometries and forces on the training structures calculated using DFT with and without SOC were compared. The results show that while SOC affects the optimized defect geometries of iodide vacancies, especially \ce{V_{I}^{-}} \cite{meggiolaro2018first}, it has no significant influence on forces, and hence, it was not included for training the force fields to limit the computational cost. These comparisons are given in SI note 4.} The detailed training procedure for the defective systems can be found in SI Note 3.

{ As a simple, straightforward test on} the accuracy of the force fields, we compare the defect migration barrier calculated with the MLFFs and with DFT, using the climbing image nudged elastic band (CI-NEB) technique \cite{henkelman2000climbing}; the results are shown in Figure~\ref{fig:figure1}c-h. The MLFF migration barriers are generally in good agreement with the DFT results,{ with differences on the scale of 0.1 eV or less. The exception is \ce{V_{I}^{0}} where the difference is \SI{0.22}{eV}. The DFT calculated trends in migration barriers for iodide interstitials ($E_{\mathrm{b}}$(\ce{I_{I}^{+}}) $>$ $E_{\mathrm{b}}$(\ce{I_{I}^{-}}) $>$ $E_{\mathrm{b}}$(\ce{I_{I}^{0}})) and iodide vacancies ($E_{\mathrm{b}}$(\ce{V_{I}^{-}}) $>$ $E_{\mathrm{b}}$(\ce{V_{I}^{0}}) $>$ $E_{\mathrm{b}}$(\ce{V_{I}^{+}})) are well captured by MLFF.} Details of the CI-NEB calculations along with all calculated migration barriers are given in SI Note 6.

In addition to energies, we also check the accuracy and transferability of these models in predicting forces. We sample structures from MD simulations at \SI{600}{K} performed using { $6\times6\times6$ cubic (216 units of \ce{CsPbI_{3}}) supercells} for each defect system and compare the forces calculated using MLFFs with those calculated using DFT. The results of these comparisons give a {$R^{2}>0.94$} and a mean absolute error {(MAE) $\leq$ \SI{54.83}{meV/\angstrom}}. For forces acting on the atoms close to the defect environment, all MLFFs have {$R^{2}>0.93$ and MAE $\leq$ \SI{61.88}{meV/\angstrom}}, illustrating the high accuracy and transferability of the MLFFs { between differently sized supercells}. The details of these MD runs, the procedure for identifying the defect environments, and the comparison of forces for all MLFFs with DFT can be found in {SI Note 6}.

Following training and validation, the force fields are used to perform { at least five independent \SI{2}{ns} long MD simulations per temperature at five temperatures between 500 K and 600 K. The MD timestep is \SI{2}{fs} and substantial cubic supercells ($\mathrm{6\times6\times6}$, 216 units of \ce{CsPbI_{3}}) are used containing only one iodide point defect to minimize interactions between the defect and its periodic images.} The volume is kept constant during each run with the lattice parameters at the different temperatures extracted from the constant temperature MD runs on pristine \ce{CsPbI3} ({see SI Note 4}). The temperature range is chosen to ensure a sufficient number of defect migration events within a lattice that is subjected to moderate temperature fluctuations. { From the mean squared displacement (MSD) curves decomposed to the chemical species, plotted in SI figure S11, one can deduce that iodide is the only species that migrates during the simulations.} Indeed, our simulations show multiple migration events for all defects, except for $\mathrm{V_{I}^{-}}$, where no defect migration was observed at all. The details of the simulation runs are given in {SI Note 7}.

\begin{figure}[htbp!]
    \includegraphics{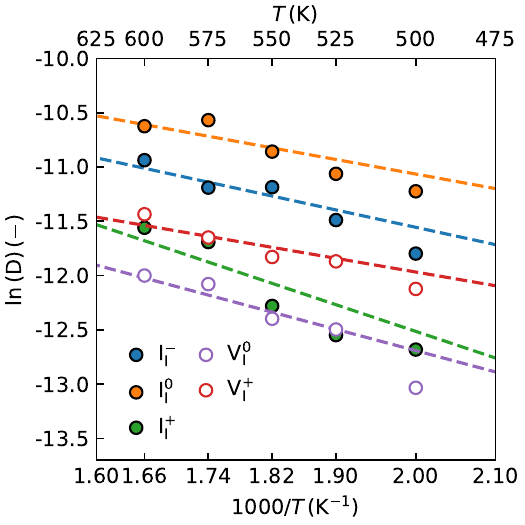}
    \caption{{Temperature dependent diffusion coefficients of halide point defects obtained from the MD simulations with the MLFFs. The filled symbols represent halide interstitials, and the open symbols represent halide vacancies. The dashed lines represent the fits to an Arrhenius expression.}}
    \label{fig:figure2}
\end{figure}

The diffusion behavior for all five mobile species can be fitted by an Arrhenius relation
\begin{equation}
    D = D_{0}\exp\left(-\frac{E_\mathrm{a}}{k_\mathrm{B}T}\right)
    \label{eqn:equation1}
\end{equation}
where $k_\mathrm{B}$ is the Boltzmann constant, $E_\mathrm{a}$ the activation energy, and $D_{0}$ the pre-exponental factor. 
The fits are shown in Figure~\ref{fig:figure2}, and the parameters extracted from the Arrhenius fit are given in Table~\ref{tab:table1}. A full explanation of how the diffusion coefficients are calculated can be found in {SI Note 7}.

\begin{table*}
    \centering
    \begin{tabular}{|c c c c|}
     \hline
     System & ${E_\mathrm{a}\,\,\mathrm{(eV)}}$ & ${D_{0}\,\,(\times10^{-3}\,\mathrm{cm}^{2}\mathrm{s}^{-1})}$ & ${D_{300K}\,\,(\times10^{-7}\,\mathrm{cm}^{2}\mathrm{s}^{-1})}$\\
     \hline
     $\mathrm{I_{I}^{-}}$ & {$\mathrm{0.21\pm0.03}$} & {$\mathrm{1.08\pm0.77}$} & {3.20}\\
     $\mathrm{I_{I}^{0}}$ &
     {$\mathrm{0.17\pm0.02}$} & {$\mathrm{0.80\pm0.41}$} &
     {11.14}\\
     $\mathrm{I_{I}^{+}}$ &
     {$\mathrm{0.30\pm0.05}$} & {$\mathrm{3.45\pm3.37}$} &
     {0.31}\\
     $\mathrm{V_{I}^{0}}$ &
     {$\mathrm{0.26\pm0.03}$} & {$\mathrm{1.17\pm0.91}$} &
     {0.50}\\
     $\mathrm{V_{I}^{+}}$ &
     {$\mathrm{0.16\pm0.03}$} & {$\mathrm{0.24\pm0.19}$} &
     {4.92}\\
     \hline
    \end{tabular}
    \caption{\textmd{Activation energies (${E_\mathrm{a}}$) and pre-exponential factors (${D_{0}}$) extracted from the Arrhenius fits, and extrapolated diffusion constants (${D_{300\mathrm{K}}}$) at room temperature.}}
    \label{tab:table1}
\end{table*}

Activation energies for diffusion ${E_\mathrm{a}}$ range {from 0.16 to 0.30 eV}, and are quite dependent on the charge state of the defects. Starting from the iodide interstitial in its most stable charge state, $\mathrm{I_{I}^{-}}$ \cite{meggiolaro2018,Xue2022}, the activation energy decreases by {0.04 eV} { for the neutral interstitial}, $\mathrm{I_{I}^{0}}$, but it increases by {0.11 eV} { for the positively charged interstitial}, $\mathrm{I_{I}^{+}}$. The iodide vacancy in its most stable charge state, $\mathrm{V_{I}^{+}}$, has the lowest activation energy for diffusion of all cases considered here. The activation energy {increases by \SI{0.10}{eV} for the neutral} vacancy, $\mathrm{V_{I}^{0}}$. For the negatively charged vacancy, $\mathrm{V_{I}^{-}}$, we have not observed any diffusion in our simulations, so it is safe to assume that in that case, the activation energy is even higher.

Note that the qualitative trend in activation energies as function of charge state obtained from the NEB calculations, Figure~\ref{fig:figure1}, is similar to that obtained from the MD simulations. Quantitatively, however, the NEB values can easily be off by a factor of two, which confirms the notion that NEB calculations might be less suitable for soft materials such as metal halide perovskites, where a representative sampling of migration paths and barriers is difficult to find among the many degrees of freedom.  

The pre-exponential factors, $D_0$ in Table~\ref{tab:table1}, also span quite a wide range of several orders of magnitude. This will be discussed in more detail below. We can use the Arrhenius expression to extrapolate the diffusion coefficients to a lower temperature, at which the infrequency of diffusion events would prohibit a direct simulation. Extrapolated diffusion coefficients (${D_{300\mathrm{K}}}$) for room temperature T = \SI{300}{K} are given in Table~\ref{tab:table1}. At this temperature, interstitials and vacancies in their most stable charge states, $\mathrm{I_{I}^{-}}$ and $\mathrm{V_{I}^{+}}$, have similar diffusion coefficients ({3.20 vs. 4.92 $\times10^{-7}\,\mathrm{cm^{2}s^{-1}}$}).

At room temperature, the neutral defects, with $\mathrm{I_{I}^{-}}$ capturing a hole or $\mathrm{V_{I}^{+}}$ capturing an electron, behave oppositely. The iodide interstitial $\mathrm{I_{I}^{0}}$ migrates faster than $\mathrm{I_{I}^{-}}$, while the iodide vacancy $\mathrm{V_{I}^{0}}$ migrates at least one order of magnitude slower than $\mathrm{V_{I}^{+}}$. Finally, $\mathrm{I_{I}^{-}}$ capturing two holes makes the interstitial $\mathrm{I_{I}^{+}}$ much less mobile, whereas $\mathrm{V_{I}^{+}}$ capturing two electrons makes the vacancy $\mathrm{V_{I}^{-}}$ immobile in the present simulations.

In a simple model of diffusion of a defect as a random walk on a lattice, the pre-exponential factor can be expressed as
\begin{equation}
    D_{0} = \frac{d^2}{z}\nu_0,
    \label{eqn:equation2}
\end{equation}
where $d$ is the step size (distance between nearest neighbor defect equilibrium positions), $z$ is the number of possible jump directions from one site, and $\nu_0$ is the attempt frequency. As an example, for an iodide vacancy $\mathrm{V_{I}^{+}}$, $d\approx 4.5$ \AA, $z=8$, and from Table~\ref{tab:table1} $D_0 = 10^{-4}$ cm$^2$s$^{-1}$, would give an attempt frequency $\nu_0 = 0.4$ THz, which indeed is a typical lattice vibration frequency in CsPbI$_3$. We conclude that the pre-exponential factors, extracted from the MD simulations, and listed in Table~\ref{tab:table1}, are within expected physical orders of magnitude.

As for the activation energies, the values listed in Table~\ref{tab:table1} are typically smaller than values found from NEB calculations, Figure~\ref{fig:figure1}. As discussed { in the introduction}, NEB calculations tend to find upper bounds, which in particular for soft materials can be quite far from the actual values{, and is likely to present a problem for TST in these materials. For example, a transition state theory (TST) study on iodide vacancy diffusion in CsPbI$_3$ found a value of 0.34 eV for the diffusion barrier of $\mathrm{V_{I}^{+}}$ \cite{Woo2022factors}. This value should be close to our NEB value of 0.44 eV, Figure \ref{fig:figure1}h and SI Table S7, the difference being explained by differences in the structures along the selected diffusion path and the exchange-correlation functional used. Both these values are considerably higher that the $\mathrm{V_{I}^{+}}$ diffusion barrier of $0.16\pm0.03$ eV found in MD, Table \ref{tab:table1}. Other TST/NEB calculations have focused on MAPbI$_3$, finding values for the diffusion barrier of iodide vacancies of 0.08 eV \cite{azpiroz2015defect},  0.26 eV \cite{yang2016fast}, 0.32 eV \cite{haruyama2015first} and 0.58 eV \cite{Eames2015ionic}. The spread in values partly comes from the different exchange-correlation functionals used \cite{Xue2021first}, but also reflects the intrinsic difficulty of properly sampling diffusion paths. Similar spreads might be expected in TST calculated diffusion barriers of other defects.}

{ Comparison to experimentally obtained diffusion barriers is also not so straightforward, which is related to the difficulties of extracting this parameter from experiments, or even identifying the microscopic nature of the diffusing species, as discussed in ref. \cite{senocrate2019solid}. Not questioning the identifications, values of 0.15-0.20 eV \cite{Reichert2020} and $0.29\pm 0.06$ eV \cite{futscher2019quantification} have been reported for iodide interstitials in MAPbI$_3$, and 0.20-0.36 eV \cite{tammireddy2021temperature}, $0.40\pm0.01$ eV \cite{schmidt2024consistent} and 0.60-0.68 eV \cite{Eames2015ionic} for iodide vacancies.} 

A closer examination of the activation energies and pre-exponential factors in Table~\ref{tab:table1} reveals a relationship, where a larger activation barrier corresponds to a larger pre-exponential factor (figure given in SI Note 8). From Equation~\ref{eqn:equation1} this implies that a change in the latter partly compensates for a change in the former, such that the diffusion coefficient is less affected by these changes. {As an example, whereas the activation barrier for $\mathrm{I_{I}^{-}}$ migration is \SI{0.05}{eV} higher than that for $\mathrm{V_{I}^{+}}$ migration, see Table~\ref{tab:table1}, the pre-exponential factor is almost five times larger.} This results in the diffusion coefficients at room temperature ${D_{300\mathrm{K}}}$ for $\mathrm{I_{I}^{-}}$ and $\mathrm{V_{I}^{-}}$ being of the same order of magnitude.   

Such a (partial) compensation between changes in the activation barrier and in the pre-exponential factor was also reported in experiments by in ref. \cite{Reichert2020} for \ce{MAPbI_{3}}, characterizing the migration of MA and I related defects, and categorizing this compensation under the Meyer-Neldel rule. That rule relates the pre-exponential factor of a reaction (or diffusion) rate to the entropy of the transition state, and states that an increase in the energy of the transition state (the activation energy) is accompanied by an increase in its entropy. The two increases then (partially) compensate one another in their effect on the reaction (or diffusion) rate, see {SI Note 8} \cite{meyer1937,du2022optimizing,Takamure2022}.

\begin{figure*}[htbp!]
    \includegraphics{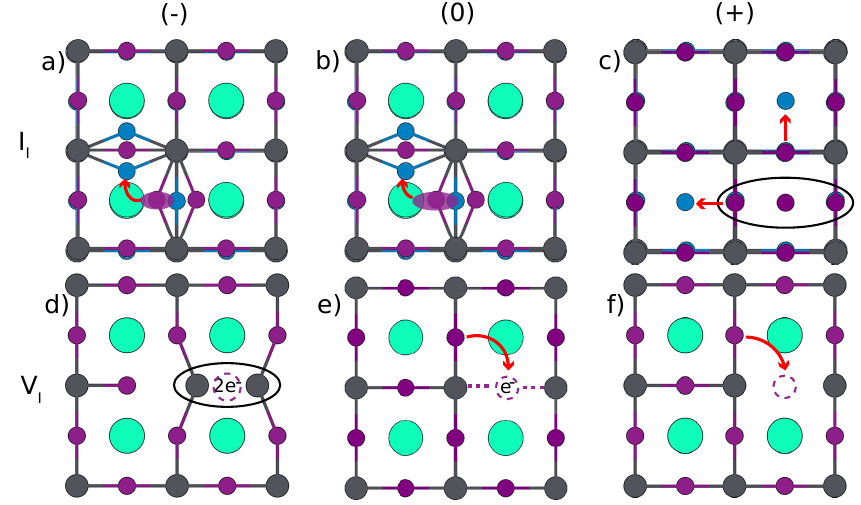}
    \caption{Schematic representations of the diffusion paths of iodide interstitials (a-c) and vacancies (d-f) in their different charge states. The arrows represent the migration directions of the ions.}
    \label{fig:figure3}
\end{figure*}

To help analyze the trends in the diffusion rates, we analyzed the structural geometries along the migration paths, depicted schematically in Figure~\ref{fig:figure3}. The iodide interstitial $\mathrm{I_{I}^{-}}$ in its most stable configuration appears in a characteristic structure, where it doubles the bridge between two Pb ions formed by a lattice iodide, as in Figure~\ref{fig:figure3}a \cite{meggiolaro2018,meggiolaro2018first,Xue2021first,Xue2022}. The typical migration path for such interstitials then consists of hopping moves of a \ce{I} atom from one \ce{Pb-I, I-Pb} bridge 
to a neighboring \ce{Pb-I-Pb} bond to form a double bridge there. The neutral interstitial $\mathrm{I_{I}^{0}}$ behaves in a similar way, but its bonding to the Pb atoms is less rigid ({SI note 9}), resulting in a slightly lower activation energy and a higher migration rate for $\mathrm{I_{I}^{0}}$. 
 
In contrast, the equilibrium bonding configuration of the positively charged interstitial $\mathrm{I_{I}^{+}}$ is quite different. It is not directly bonded to Pb atoms, but instead to two lattice iodides, and forms a linear iodide trimer \ce{I3}, see Figure~\ref{fig:figure3}c, \cite{meggiolaro2018,meggiolaro2018first,Xue2021first,Xue2022}.
The diffusion of this interstitial leads to a migration path where the interstitial kicks out a lattice iodide, which then becomes the new interstitial. This migration path has a higher activation energy, resulting in a considerably lower migration rate for $\mathrm{I_{I}^{+}}$ compared to the other two charge states of iodide interstitials. 

In the case of iodide vacancy, Figure~\ref{fig:figure3}d-f, the migration is characterized by an exchange of an \ce{I} vacancy with an \ce{I} atom, involving a $90^\circ$ rotation around the \ce{Pb-I} bond. The neutral vacancy $\mathrm{V_{I}^{0}}$ has a similar migration path, but as a result of $\mathrm{V_{I}^{+}}$ capturing an electron, the vacancy is more strongly bonded to the two positively charged \ce{Pb} surrounding it. Thus, $\mathrm{V_{I}^{0}}$ experiences a higher activation barrier than $\mathrm{V_{I}^{+}}$. This effect is even more significant for $\mathrm{V_{I}^{-}}$ where after adding another electron the bond to the surrounding Pb ions becomes even stronger, { which is consistent with a shortening of the \ce{Pb-Pb} distance, see Tables S3 and S5 in the SI. Indeed}, no migration was observed for $\mathrm{V_{I}^{-}}$ in the simulated temperature range. 

\section{Discussion and Conclusions}
A fundamental problem is extracting the nature of a diffusing defect from experiment. Some studies focus on iodide vacancies \cite{senocrate2017nature,senocrate2019solid,tammireddy2021temperature}, an identification that is typically based upon varying the chemical potential of iodide from iodine poor to rich, and observing the ion conductivity decrease, which is then interpreted as a decrease of iodide vacancy concentration. However, for perovskites, with their charged iodide defects, this approach may be troublesome. A change in the iodine chemical potential is compensated under intrinsic conditions by a change in Fermi level. As a result of this the defect formation energies, and therefore their concentrations, are not changed, as explained in ref. \cite{Xue2022}. 

First-principles calculations often find that the iodide interstitial is more easily formed than the vacancy \cite{meggiolaro2018,meggiolaro2018first,Xue2021first,Xue2022}, and should be present in a larger concentration under equilibrium conditions, except under very iodine poor conditions. As an alternative explanation for the decrease in ion conductivity going from iodine poor to rich, we suggest that the intrinsic Fermi level decreases \cite{Xue2022}, where at some point the iodide interstitials change their charge state from $\mathrm{I_{I}^{-}}$ to $\mathrm{I_{I}^{+}}$. The latter are much less mobile according to Table~\ref{tab:table1}, which should lead to a lower iodide conductivity. Some experimental studies indeed focus on the role of iodide interstitials \cite{futscher2019quantification,Reichert2020,Ni2022} Interestingly, the value $0.29\pm 0.06$ eV found in ref. \cite{futscher2019quantification} for the activation energy of the diffusion attributed to the iodide interstitial $\mathrm{I_{I}^{-}}$, overlaps with the value {$0.21\pm 0.03$ eV} found from the MD simulations, see Table~\ref{tab:table1}.

{ In ref. \cite{Samatov2024}, atomic migration is studied at grain boundaries (GB) in \ce{CsPbBr3} using MD simulations based on MLFFs, where it is found that also there the halide atoms (Br in this case) form the only migrating species. Assuming that diffusion processes in \ce{CsPbBr3} and \ce{CsPbI3} are similar, the main difference seems to be that diffusion barriers at grain boundaries seem to be extremely low, even with respect to the low barriers found in the present case, see Table \ref{tab:table1}.}

In conclusion, we trained machine learned force fields for different charge states of iodide interstitial and vacancy defects in \ce{CsPbI3}. Using these force fields we performed long-timescale MD simulations to study the temperature-dependent diffusion behavior of these defects. Our simulations suggest that out of the six investigated species (positive, negative, and neutral interstitials and vacancies), five are mobile. When closely comparing these five mobile species, we found that iodide interstitials and vacancies in their most stable charge states, as they occur under (near) equilibrium conditions ($\mathrm{I_{I}^{-}}$, $\mathrm{V_{I}^{+}}$), migrate at similar rates at room temperature. Neutral iodide interstitials are somewhat faster, but neutral vacancies are one order of magnitude slower. Oppositely charged interstitials $\mathrm{I_{I}^{+}}$, such as they can occur in device operating or reverse bias conditions are significantly slower, and the oppositely charged vacancy $\mathrm{V_{I}^{-}}$ can be considered as relatively immobile.

Overall, our findings indicate that defect migration rates in halide perovskites undergo significant changes upon charge capture during device operation conditions. In particular, we highlight the role of iodide interstitials, not only due to their rapid migration kinetics but also their high abundance, driven by favorable thermodynamic conditions for their formation. Moreover, their ability to capture charges alters their mobility and can trigger redox reactions. These processes are critical to consider when interpreting macroscopic observations, such as ion conductivity in perovskite films and the evolution of I-V curves during device operation. Our work also paves the way for further studies on the complex interplay and reactions between different defect types and charges in halide perovskites.

\begin{acknowledgement}
The authors thank Henry Kwan for testing the parameters for CI-NEB calculations. V.T. and S.T. acknowledge funding from Vidi (project no. VI.Vid.213.091) from the Dutch Research Council (NWO).
\end{acknowledgement}

\begin{suppinfo}
Supporting information will be made available on publication.
\end{suppinfo}

\textbf{Conflict of Interest} \par
The authors declare no conflict of interest.

\bibliography{references}

\end{document}